\documentclass[preprint,groupedaddress,preprintnumbers,amsmath,amssymb]{revtex4}
\usepackage{epsfig,longtable,color}

\usepackage{graphicx}
\usepackage{dcolumn,color} 
\usepackage{bm}
\usepackage{times}

\begin{document} 

\title{Hydrophobic Interactions and Dewetting between Plates with Hydrophobic and Hydrophilic Domains\footnote{In Press, J. Phys. Chem. C}}
\author{Lan Hua, Ronen Zangi\footnote{Present address: Department of
   Organic Chemistry I, University of the Basque Country, Avenida de
   Tolosa 72, 20018 San Sebastian, Spain}, and B. J. Berne}

\affiliation{Department of Chemistry, Columbia University, 3000 Broadway, New York, NY 10027.}

\lineskip=8pt minus 1 pt \lineskiplimit=7pt

\date{\today}


\begin{abstract}
  We study by molecular dynamics simulations the wetting/dewetting
  transition and the dependence of the free energy on distance between
  plates that contain both hydrophobic and hydrophilic particles.  We
  show that dewetting and strength of
  hydrophobic interaction is very sensitive to the distribution of
  hydrophobic and hydrophilic domains. In particular, we find that
  plates characterized by a large domain of hydrophobic sites induce a
  dewetting transition and an attractive solvent-induced
  interaction. On the other hand, a homogeneous distribution of the
  hydrophobic and hydrophilic particles on the plates prevents the
  dewetting transition and produces a repulsive solvent-induced
  interaction. We also present results for a kind of ``Janus
  interface'' in which one plate consists of hydrophobic particles and
  the other of hydrophilic particles showing that the inter-plate gap
  remains wet until steric constraints at small separations eject the
  water molecules. Our results indicate that the Cassie equation, for
  the contact angle of a heterogeneous plate, can not be used to
  predict the critical distance of dewetting. These results indicate
  that hydrophobic interactions between nanoscale surfaces with strong
  large length-scale hydrophobicity can be highly cooperative and thus
  they argue against additivity of the hydrophobic interactions
  between different surface domains in these cases. These findings are
  pertinent to certain protein-protein interactions where additivity
  is commonly assumed.

\end{abstract}

\maketitle

\section{Introduction}
\label{sec:Introduction}

The hydrophobic interaction is one of the major driving forces in
various self-assembly processes such as protein folding, the formation
of membranes and micelles, molecular recognition and surfactant
aggregation.\cite{ball_2003,Scheraga_jbsd98,Fersht-folding,pratt3,chandler_nature05}
The nature of the hydrophobic effect is length-scale dependent.  While
hydrophobicity at small length-scales is associated with small
distortions of the hydrogen bond connectivity between the water
molecules, large-scale hydrophobicity is driven by a significant
disruption of the hydrogen bonds network of interfacial
waters.~\cite{stillinger,berne1,Lee_jcp84,LCK_jpcb99,bratko,Berne3_2003,thirumalai1,Huang_pnas00,Huang_jpcb02}
The crossover from small to large scale behavior occurs when the
radius of the hydrophobe is about 1 nm. One of the signatures of
large-scale hydrophobicity is the existence of a dewetting (drying)
transition; on bringing two large hydrophobic plates from far apart
into contact, a critical distance exists below which the water
confined between the two plates is unstable in its liquid state and
evaporates. The dewetting transition has been predicted
theoretically\cite{chandler4_1998,LCK_jpcb99,Huang_jpcb02,Berne3_2003,Ashbaugh_rmp06}
and observed in computer
simulations\cite{Lum97,Wallqvist_jpc95,Bolhuis_2000,Wolde_pnas02,chandler_jpcb03,Berne3_2003,hummer,Li_jacs06,Ronen_jacs07,Hua_jpcb07,Liu_nature05,chandler_pnas07,chandler_jpcb07,Chandler_jpcb08}.
In a recent experiment using interfacial-force
microscopy\cite{sandia06}, a nanoscale bubble was directly observed
between superhydrophobic surfaces. Nevertheless, the drying transition
is sensitive to the magnitude of the solute-solvent interaction.  This
was demonstrated by showing that a small change of the attraction
between the wall of a nonpolar carbon nanotube and water can induce a
wetting transition\cite{hummer}.  In this case, the transition point
is characterized by a two-state model; in one state the channel is
filled with water and in the other it is empty\cite{chandler_jpcb03}.
As the solute-solvent attractive interaction increases, the dewetting
transition becomes less pronounced and eventually
disappears.\cite{Huang_jpcb02,Berne3_2003,Zhou_science04,chandler_jpcb07}
In some cases the kinetic pathway for dewetting in the cavitation
transition has been studied using the powerful transition path
sampling method\cite{Bolhuis_2000} and the string
method\cite{chandler_pnas07}.  The hydrophobicity of a
surface is related to the value of the contact angle of a water
droplet on that surface. Based on Young's equation, a surface with a
contact angle larger than 90$^\circ$ is defined as hydrophobic. It is
interesting to point out that, a hydroxylated silica surface with its
partial atomic charges scaled by a factor of less than 0.4 was found
to be macroscopically hydrophobic\cite{Rossky_jpcb07}.  In addition,
it has been argued that the mechanism for attractive mean forces
between the plates is very different depending on the nature of the
solute-solvent interaction\cite{Choudhury05}.

Both morphology and structure of the interacting hydrophobic surfaces
are important to the existence and kinetics of
dewetting.~\cite{Luzar_jcp00,Christ01,Rossky_jpcb07,chandler_fd08}
In general, the surface of proteins is non-homogeneous with respect to
hydrophobicity, roughness and topology. For example, simulations
indicate that protein surfaces can be characterized by hydrophobic
regions that are heterogeneously ``small'' or ``large'' based on the
length-scale dependence of hydrophobic
hydration.\cite{Cheng_nature98}. Thus, it is very challenging to
predict the existence of the drying transition and determine its role
in protein folding from the properties of the protein interface
alone. In a recent study, we found that a simple hydrophobic scoring
function, based on aligned hydrophobic surface areas, is not
sufficient to predict whether the assembly of protein units will
exhibit dewetting or not.~\cite{Hua_jpcb07} Further improvement of the
scoring function should, therefore, include contributions from factors
which were not taken into account, such as surface roughness and
heterogeneity.

A macroscopic theory relates the critical distance of the
dewetting transition to the value of the contact angle of a water
droplet on the surface,~\cite{Berne3_2003}
\begin{equation}
\label{eq:macroscopic_theory} D_c
=\frac{-2\gamma_{lv}\cos{\theta_c}}{(P-P_v) + 2\gamma_{lv}/R_m}
\end{equation}
where $\gamma_{lv}$ is the liquid/vapor surface tension for water,
$\theta_c$ is the contact angle of water on the hydrophobic plate,
$P$ is the pressure of the liquid (water), $P_v$ is the vapor
pressure of water, and $R_m$ is the radius of the plates. For
plates of nanoscale size, the term in the denominator $(P-P_v)$ is
much smaller than $2\gamma_{lv}/R_m$ and can be ignored, yielding,
\begin{equation}
D_c \approx -R_m \cos{\theta_c}
\end{equation}
Thus, a knowledge of the contact angle of a heterogeneous surface
is enough to predict the critical distance for dewetting. The
contact angle $\theta$ on a heterogeneous solid surface can be
approximately predicted by the Cassie
equation\cite{Cassie_dfs48,Cassie_dfs52}. For a two-component
surface,
\begin{equation}
\label{eq:cassie}
\cos\theta = f_1\cos\theta_1 + f_2\cos\theta_2
\end{equation}
where, $f_i$ is the fractional area of the surface with a contact
angle of $\theta_i$. Simulations of droplets on heterogeneous
patterned surfaces indicate that the Cassie equation holds for
domains that are sufficiently small relative to the droplet. On
the other hand, when the size of the heterogeneous domains is much
larger than that of the droplet, there is a breakdown of the
Cassie equation.~\cite{Lundgren_07,Brandon_03,Padday_78} In the
current work, we are interested in cases in which the size of the
domains is small compared with the size of a typical macroscopic
droplet. For example: under what conditions is the Cassie equation
obeyed?; and how does the distribution of hydrophobic and
hydrophilic domains (in terms of size and shape) determine the
averaged contact angle $\theta$ for
a given $f_i$? 

Recent studies on hydrophobic hydration and hydrophobic interactions
between hybrid polar/nonpolar
nanoassemblies\cite{Koishi_prl04,Koishi_jcp05,Rossky_jpcc07} show that
the pattern of nonpolar-site distribution is very important for the
dewetting transitions to occur\cite{Koishi_jcp05}.  Hydrophilic
borders surrounding a nanoscale hydrophobic patch reduce considerably
the patch's ability to repel water from the first hydration layer. In
fact, even a single hydrophilic site at the center of the surfaces
prevents complete drying of the confined region\cite{Rossky_jpcc07}.
In addition, it was found that an increase in the pressure of the bulk
water blurs the difference between interfacial water density next to
hydrophilic and hydrophobic surfaces\cite{Rossky_jpcc07}.  However,
none of these studies quantitatively addressed the correlation between
the distribution of hydrophobic sites and dewetting as well as the
strength of hydrophobic interactions. The related problem of how water
behaves next to single heterogeneous surfaces has been addressed in a
recent study using coarse grained modeling of the interace between
water and heterogeneous surfaces\cite{chandler_fd08}.

It is also of great interest to investigate the extreme case where one
plate is highly hydrophobic and the other hydrophilic. In an intriguing
experiment Granick and coworkers investigated just such a system. They
studied a Janus interface in which a water slab is trapped between a
hydrophobic wall on one side and a hydrophilic wall on the
other\cite{granick1_2002} and found that it prevents any macroscopic
drying or cavitation of the liquid, which in any case would be
strongly affected even by relatively weak van der Waals forces.

Much work has been done on the non-additivity of hydrophobic
interactions through calculations of PMF between simple small
hydrophobic solutes in dilute
solution.~\cite{Rank_ps97,Czaplewski_ps00,Czaplewski_jcp02,Czaplewski_ijqc02,Czaplewski_bc03,Shimizu_jcp01,Shimizu_pro02,Chan_jacs05,Palma_cpl96,Palma_ebj98,Levitt_ps97,Levitt_pnas01,Ghosh_jpcb03,Hummer_jacs99}
It has been suggested that long range hydrophobic interactions
caused by the many-body character of the PMF~\cite{Martorana_bpj97}
is relevant for the energetics
within~\cite{Moza_pnas06,Keskin_jmb05,Reichmann_pnas05} or
between~\cite{Moza_pnas06} distinct ``hot regions'' of interacting
protein surfaces. This is very important for the prediction and
modulation of protein-protein interactions. Although the
non-additive effect was found to be insensitive to the strength of
the solute-solvent van der Waals
interaction~\cite{Czaplewski_ijqc02}, it was shown to increase with
the size of the hydrophobic
solutes~\cite{Ghosh_jpcb03,Czaplewski_ijqc02}. The neglect of the
length-scale dependence of hydrophobic solvation results in the
failure to predict cooperativity for three-body hydrophobic
association in current surface area based nonpolar
models~\cite{Chen_pccp08}. Previous studies of non-additivity were
performed for simple hydrophobic solutes like methane, where the
sizes of the clusters that form are small. Since proteins surfaces
are heterogeneous with mixed ``small'' and ``large'' hydrophobic
regions, it is of interest to study the non-additive effect in the
association of nanoscale hydrophobic assemblies in a heterogeneous
context.

Our aim in this paper is to determine how the distribution of the
constituent hydrophobic and hydrophilic particles determines the
hydrophobic interaction between such amphiphilic surfaces. We perform
systematic molecular dynamics simulations of different surfaces
characterized by the same size and same number of hydrophobic and
hydrophilic particles, which are distributed in different
patterns. Our results show that the behavior of water confined between
two identical amphiphilic plates greatly depends on the distribution
of the hydrophobic and hydrophilic particles on the plates. This
effect is manifested by the existence or absence of a dewetting
transition as well as by attractive or repulsive solvent induced
interactions between the plates. We propose a parameter, the average
number of hydrophobic nearest neighbors of a hydrophobic particle, to
describe the degree of clustering of the hydrophobic particles on the
surface, and show that the behavior of confined water (dewetting or
cavitation) is correlated with the value of this parameter. We also
observe that for drying to occur the minimum area for clustered
hydrophobic particles on the surface is (1.04$\times$1.04) nm$^2$ and
we call this the critical area for the drying transition. These
results demonstrate a strong cooperativity in the hydrophobic
interaction between surface hydrophobic domains of varied length
scale. In addition, our results show that it is not possible to
predict the critical distance for dewetting between heterogeneous
surfaces by using the contact angle obtained by the Cassie equation
(Eq.(\ref{eq:cassie})) in Eq.(\ref{eq:macroscopic_theory}).
We also carried out simulations of water confined to a Janus interface
(where one plate is hydrophobic and the other is hydrophilic).
In these simulations we find that the gap between these plates remains
wet if the partial charges on the hydrophilic plates are sufficiently
large. However, for small partial charges a drying transition is observed.
In this case, the critical distance for dewetting can be obtained from the average of the critical distance for the pure hydrophobic system (where the two plates are hydrophobic) and that for the pure hydrophilic system.

\section{Methods}
\label{sec:methods}

We studied the thermodynamics of the association process of two
identical large amphiphilic plates composed of hydrophobic and
hydrophilic particles. Each surface is represented by, a
single-layered plate of, 49 atoms arranged in a 7$\times$7 square
lattice with a bond length of 0.32 nm. Each plate has 25 hydrophobic
atoms and 24 hydrophilic atoms. The Lennard-Jones (LJ) parameters for
the interaction between plate atoms on different plates are
$\sigma_{plt}=0.40$ nm and $\epsilon_{plt}=0.50$ kJ/mol, values lying
in the same range as in our previous work.\cite{Zangi06,Ronen_jacs07}
We represent hydrophilic atoms in two ways. In the first approach, the
hydrophilic particles are represented by the same LJ parameters as for
hydrophobic particles, however, they have nonzero partial
charges. Since all the surfaces we generated are electrically neutral,
the number of positive and negative particles is the same. In one set
of simulations we studied hydrophilic particles with partial charges
of Q=$\pm 0.4$ \emph{e}, and in another set with partial charges of
Q=$\pm 0.8$ \emph{e}. In the second approach, we represented
hydrophilic particles with a large well-depth of the LJ potential
($\epsilon^{phil}_{plt}=$1.30--2.0 kJ/mol), which is considerably
larger than that for the hydrophobic particles
($\epsilon^{phob}_{plt}=$0.5--1.0 kJ/mol). Nevertheless, the LJ
diameter was taken to be the same as that for the hydrophobic
particles, $\sigma_{plt}=0.40$ nm. We investigated five different
patterns of hydrophobic/hydrophilic particle distributions on the
plates (pattern I-V) (see Figure~\ref{fig:hetero_1}(a)).

The two plates were solvated in 1147 water molecules. We chose the
SPC/E model\cite{Berendsen87} of water and used combination rules
(arithmetic average for $\sigma$ and geometric average for $\epsilon$)
to calculate the water-plate interactions. Analysis of the water
density profile next to the plates suggests that the strength of the
attraction between the water molecules and these plates is very
similar to that of water and a hydrocarbon monolayer described at
atomic level\cite{Li06}.
In addition, recent simulations\cite{Zangi08} investigating the contact
angle of water on the most hydrophobic surface studied here,
$\epsilon^{phob}_{plt}=$0.5 kJ/mol, find its value to be 119$^\circ$.
This value is similar to the values obtained for self-assembled
octadecanethiol monolayer on silver\cite{Zhu02} (117$^\circ$), and on
gold\cite{Plant95} (105$^\circ$) surfaces.

During simulations, the positions of the
plate atoms are held fixed, interactions between atoms on the same
plate are excluded, and the orientation of the two plates with respect
to each other is parallel and in-registry. The alignment of the
surface patterns with respect to each other is such that hydrophobic
particles on one plate are superimposed on hydrophobic particles on
the other plate. For hydrophilic particles, positive charged particles
on one plate are superimposed on negative charged particles on the
other plate.

We used the Molecular Dynamics (MD) package GROMACS version
3.1.4\cite{Lindahl01b} to perform the computer simulations, with a
time step of 0.002 ps. The bond distances and angle of the water
molecules were constrained using the SETTLE
algorithm\cite{Miyamoto92}. The system was maintained at a
constant temperature of 300 K and pressure of 1.0 bar using the
Berendsen thermostat\cite{Berendsen-thermo}. The electrostatic
forces were evaluated by the Particle-Mesh Ewald method (with grid
spacing of 0.12 nm and quadratic interpolation) and the LJ forces
by a cutoff of 1.0 nm.

The potential of mean force (PMF) between the two plates was
computed from the mean force acting on each of the
plates\cite{Pangali78,Watanabe86}. Then the mean force acting
between the plates along their axis of separation was integrated as
a function of the distance between the plates, $d$, to yield the
free energy profile. As the PMF represents only relative values, it
was shifted such that the free energy of the state at the largest
separation corresponds to zero. For each pattern (I-V), we performed
46 simulations with different values of plate separation, $d$,
ranging from 0.36--2.0 nm.  At each value of $d$, the system was
equilibrated for 2.0 ns and data was collected for 4.0 ns. At points
where the force converged slowly (around the wetting/drying
transition), the data collection stage was extended for an
additional 5.0 ns. In the analysis of the hydration of the plates at
each $d$ for each pattern, the error in the quantities obtained from
the simulations was estimated using the block averaging
method\cite{Flyvbjerg89}.

\section{Results}
\label{sec:results}

\subsection*{The dependence of solvation on the distribution of
hydrophobic particles}

The behavior of water molecules between two amphiphilic plates depends
on the distance between the plates, the distribution of hydrophobic
particles on the surface, and the nature and strength of the
interaction between the particles (hydrophilic and hydrophobic) and
water.~\cite{Zhou_science04} First we study plates in which the
hydrophilic sites are charged spheres with either $|Q|=0.4$~\emph{e}
or $|Q|=0.8$~\emph{e} to represent weak and strong hydrophilic
particles, respectively.

Figure.~\ref{fig:hetero_2} (a) shows the density of water between
two plates for the five different distribution patterns (shown in
Figure~\ref{fig:hetero_1}(a)) as a function of the inter-plate
distance when the magnitude of the partial charge of the hydrophilic
particles is equal to 0.4~\emph{e}. Dewetting transition was
observed in the inter-plate region for all the patterns except for
pattern V, which is characterized by a uniform distribution of
hydrophobic and hydrophilic particles. An absence of a drying
transition for a uniform distribution of polar and non-polar sites
was reported by Koishi et al\cite{Koishi_jcp05}. In pattern V, the
sharp decrease of water density near $d=0.72$ nm is due to steric
effects, i.e.  at $d=0.72$ nm a layer of water cannot fit between
the two plates.
Note that $d$ is the distance between the center of mass of the particles on each plate along the $z$-axis.
Thus, the available space between the plates is, approximately, $d-\sigma_{plt}$.
Interestingly, pattern IV with charged particles
arranged in the center of the plates can dewet when
$|Q|=0.4\emph{e}$. {Dewetting in this pattern might be
due to surface dipoles of small magnitude formed by charged
particles in which the distance between neighboring positive and
negative charged particles is small. This is in agreement with the
finding that a silica surface, with partial charges scaled by 0.4,
is macroscopically hydrophobic.~\cite{Rossky_jpcb07} The critical
distance, $D_c$, for dewetting for each pattern is listed on
Table~\ref{tbl:pattern_dewet}.

In order to correlate the geometrical pattern of each
plate with its critical distance, we propose a parameter, the
average number of hydrophobic nearest neighbors ($N_{nn}$) to
characterize each pattern. We define $N_{nn}$ as follows:
\begin{equation}
\label{eq:nearest_neighbor}
N_{nn}=\frac{1}{M}\sum_{i=1}^{M}N_i,
\end{equation}
Where $N_i$ is the number of nearest hydrophobic neighbors of
hydrophobic particle $i$, which has the maximum value of 4, and $M$=25
is the number of hydrophobic particles.   We assume that the closer the
hydrophobic cluster is to the center of the plate the larger will be
the critical distance $D_c$ for dewetting, but due to the small size
of the plates in our system, $D_c$ might not be sensitive to the
distance between a hydrophobic particle and the center of the
plate. Based on Equ.\ref{eq:nearest_neighbor}, we find a strong linear
correlation between $D_c$ and $N_{nn}$ with the correlation
coefficient $r$ of 0.999 (see Fig.~\ref{fig:hetero_2} (b)). Thus, as a
measure of the degree of clustering for the assembly of hydrophobic
particles on the plate, $N_{nn}$ is capable of discriminating between
the different patterns with respect to the solvation between two
plates.

When the magnitude of the charge of the hydrophilic particle is
$Q=0.8$ \emph{e}, a similar trend was found for the effect of the
distributions of hydrophobic particles on the solvation of the
inter-plate region as in the case for $Q=0.4$ \emph{e}. However, a
dewetting transition is suppressed in most patterns.  Water
depletion was found in patterns I--III and the extent of depletion
is linearly proportional to $N_{nn}$ for each pattern at the same
plate-plate separation. Patterns IV--V stay hydrated until water is
expelled because of steric effects. Note that water molecules are
trapped more tightly in pattern V than in pattern IV, leading to a
higher density of confined water than in the former, which is also
consistent with their relative values of $N_{nn}$ (see
Fig.~\ref{fig:hetero_3}). For all the patterns water density does
not
decrease to zero at $d<0.72$ nm 
as in the case of $Q=0.4$ \emph{e}, that is due to the strong
electrostatic attractive interaction between water and the charged
particles which reduces the distance for steric expulsion.

Although we did not observe strong dewetting transitions for most
patterns when $Q=0.8$ \emph{e}, we did observe stepwise cavitations
for pattern II and pattern III. These patterns are characterized by
multiple hydrophobic domains of different size distributed on the
plate surfaces. Pattern II, which has the second largest $N_{nn}$,
contains two hydrophobic domains arranged in 4$\times$4
(1.36$\times$1.36 nm$^2$) and 3$\times$3 (1.04$\times$1.04 nm$^2$)
square lattice domains. During the association of two plates, a big
cavity is first formed between the larger hydrophobic domains (see
Fig.~\ref{fig:hetero_3} (b) when $d=0.84$ nm). This is followed by the
formation of a second cavity between the smaller hydrophobic domains
at smaller plate-plate distance (see Fig.~\ref{fig:hetero_3} (b) when
$d=0.68$ nm). For $d<$0.68 nm the hydrophobic domains are dry, and the
water molecules between the hydrophilic domains are finally squeezed
out into the bulk. 
It is interesting to point out that we did not find any stepwise
cavitation for this pattern for weak hydrophilic particles ($Q=0.4$
\emph{e}). Instead, a large cavity forms without preference for the
particles' hydrophobicity, covering hydrophobic and hydrophilic
regions at the same time (see Fig.~\ref{fig:hetero_3} (b) when
$d=0.84$ nm).  The stepwise cavitation we observe for pattern II is
also observed for pattern III when $Q=0.8$ \emph{e}. However, in
this case the cavities appear at smaller plate-plate distances,
which almost can not contain more than one layer of water, compared
with that for pattern II (see Fig~\ref{fig:hetero_3} (c)). Thus,
strongly hydrophilic particles restrain cavities locally to regions
between hydrophobic domains and the cavities occur one after another
based on the size of hydrophobic domains when bringing two plates
from far apart into contact. These observed stepwise
cavitations might be related to the size dependence of the amplitude of
interfacial capillary-wave fluctuations as well as the probability
of tube formation bridging vapor-film interfaces involved in the
dewetting dynamics in the confined region.\cite{chandler4_1998} The
smallest hydrophobic domain area capable of inducing a cavity
between the two plates, is found to be 1.04$\times$1.04 nm$^2$ (see
Fig~\ref{fig:hetero_3} (d)), which is consistent with the first
cavitation in pattern III (Fig~\ref{fig:hetero_3} (c) when $d=0.68$
nm).  Comparing the distances at which each cavity is formed during
the association of the two plates for pattern II and pattern III, we
find that stepwise cavitation for each pattern occurs at inter-plate
distances which are linearly correlated with $N_{nn}$. Thus, the
larger the value of $N_{nn}$ for the entire pattern, the larger will
be the gap distances of the stepwise cavitation.

In order to investigate the sensitivity of these results to the nature
of the particle's hydrophilicity, we repeated these simulations, for
the case where the hydrophilic particles were uncharged (neutral LJ particles),
but with a larger attractive interaction, $\epsilon_{plt}$, than for hydrophobic
particles. In order to probe which values of $\epsilon_{plt}$ can be
regarded as hydrophilic and which hydrophobic, we performed
additional simulations where all of the particles were taken to be
the same and found that the threshold value of $\epsilon_{plt}$
below which a drying transition is observed is 1.0~kJ/mol. Thus, we
define particles with $\epsilon_{plt} <$ 1.0 kJ/mol as hydrophobic
and particles with $\epsilon_{plt} >$ 1.0 kJ/mol as hydrophilic.
Returning to the studies on heterogeneous surfaces, we performed
simulations with $\epsilon^{phil}_{plt}=2.0$ kJ/mol for hydrophilic
particles and $\epsilon^{phob}_{plt}=0.75$ kJ/mol for hydrophobic
particles. Fig.~\ref{fig:hetero_4} shows the results for the two
extreme cases of the distribution of hydrophobic/hydrophilic
particles. We find that when hydrophobic particles are placed at the
center of the plates as in pattern I, a strong dewetting transition
is observed in the inter-plate region (red); whereas the inter-plate
region remains hydrated if the distribution is uniform as in pattern
V (green). These results are qualitatively the same as the case
where the hydrophilic particles are represented as charged
particles.

In pattern I-V, the alignment of the two plates with respect to each
other is such that the $(X,Y)$ coordinates of the hydrophobic and
hydrophilic particles on one plate are the same as on the other
plate. It is also interesting to examine the behavior of water
between two different surfaces. We study an extreme case of two
plates which form a Janus interface \cite{granick1_2002} (see
Fig.~\ref{fig:hetero_1}(b)). One plate (plate1 in
Fig.~\ref{fig:hetero_1}(b)) is hydrophilic with positive and
negative charged particles uniformly distributed on an 8$\times$8
square lattice with the same bond length as for pattern I-V, while
the other plate (plate2) is purely hydrophobic. The same simulations
were performed for this case as for each of the other patterns. The
density of water in this Janus-interface is shown in
Fig.~\ref{fig:hetero_6}(a) with respect to different partial charges
$Q$. The Janus interface dewets when $Q=0.4$ \emph{e} and its
critical distance $D_c$ for dewetting is about 1.18 nm, which is a
little bit smaller than the one ($D_c=$1.28 nm) for $Q=0.0$ \emph{e}
in which case both plates are purely hydrophobic. However, when the
charges are increased to $Q=0.8$ \emph{e}, no dewetting is found
between the two plates. Fig.~\ref{fig:hetero_6} (b) and (c) show the
density of water along the Z-axis of the simulation box (which is
perpendicular to the plate surfaces) for inter-plate distances
$d=0.72$ nm and $d=1.24$ nm, respectively. Only water molecules
inside a rectangular box along the Z-axis of the simulation box with
$|X|<=1.1$ nm and $|Y|<=1.1$ nm (in the XY plane) are considered in
these plots. (The origin of the coordinate system is the mid-point
of the straight line connecting the centers of mass of the two plates.
The hydrophilic plate is placed at the negative values of the $z$-axis.)
These distributions indicate that in the presence of one
purely hydrophobic surface, the strong hydrophilic surface ($Q=0.8$
\emph{e}) attracts water into the gap even when it would be
geometrically impossible for this gap to accommodate one layer of
water for $Q=0.4 \mbox{ or } 0.0$ \emph{e} (see the sharp peak in
red around $z=0$ nm in Fig.~\ref{fig:hetero_6}(b)).  When $Q=0.4$
\emph{e}, the density of water near the outside surface (toward
solvent) of the weak hydrophilic plate is similar to that near
hydrophobic surface with $Q=0.0$ \emph{e} (see the peaks in black
and green with similar magnitude near $z=-0.7$ nm for $d=0.72$ nm
and those near $z=-1.0$ nm for $d=1.24$ nm). However, the density of
water in the gap region increases for $Q=0.4$ \emph{e} compared to
$Q=0.0$ \emph{e} for the plate-plate distance of 1.24 nm. These
results are consistent with the observation of dewetting in pattern
IV corresponding to $Q=0.4$ \emph{e} with hydrophilic particles
distributed at the centers of the plates and hydrophobic particles
on their borders. It also indicates that the weak hydrophilic plate
with $Q=0.4$ \emph{e} is macroscopically hydrophobic. Thus, the
behavior of water in the Janus interface depends on the polarity of
the hydrophilic surface, i.e.  the strength of the interactions between
hydrophilic particles and water.

Based on the simple macroscopic theory (see
Equ.~\ref{eq:macroscopic_theory}), the average of $\Delta \gamma$
($\Delta \gamma=-\gamma_{lv}\cos{\theta_c}$) should be used in the
calculation of $D_c$ for the two different plates in the Janus
interface. To predict $D_c$ for the Janus case with $Q=0.4$ \emph{e}
based on this simple theory, we repeated simulations for two
identical, purely hydrophilic plates with partial charge $Q=0.4$
\emph{e} (same as plate1 in Fig.~\ref{fig:hetero_1}(b)). A
dewetting transition was observed between these two plates and the
critical distance $D_c$ for dewetting was found to be in the
neighborhood of 1.10 nm. We then determined the critical distance
for the Janus case with $Q=0.4$ \emph{e} from the average of $D_c$
for the pure hydrophilic plates ($D_c=1.10$ nm) with $Q=0.4$
\emph{e} and for the pure hydrophobic plates ($D_c=1.28$ nm). The
average is 1.19 nm, essentially equal to the value ($D_c=1.18$ nm)
obtained from simulation of the Janus interface, indicating that the
critical distance of dewetting in Janus interfaces can be predicted
from the given $D_c$ for pure hydrophilic plates (same to plate1 in
Fig.~\ref{fig:hetero_1} (b)) and for pure hydrophobic plates (same
as plate2 in Fig.~\ref{fig:hetero_1} (b)).

\subsection{Potentials of mean force}

Figure~\ref{fig:hetero_5} (a) and (b) show the water induced PMF of
the amphiphilic plates for the different patterns (I-V), as a
function of the plate-plate distance.  For $Q=0.4$ \emph{e} (see
Fig.~\ref{fig:hetero_5} (a)), the water induced force between two
plates is attractive for pattern I--IV, while repulsive for pattern V,
which has the smallest value of $N_{nn}$ among all five patterns and is the
only one that does not display a dewetting transition. The
difference in the water induced PMF for bringing the two plates from
far apart to contact ($d$=0.40 nm) between the different patterns is
very large; it is about 230 kJ/mol between pattern I and pattern V.
While patterns III and IV display free energy barriers for
dewetting, patterns I and II do not. At small distances, the water
induced PMF is less negative for pattern I than for pattern II,
probably because the water molecules like to stay in the gap due to
their strong electrostatic interaction with charged particles on the
edges of plates in pattern I compared with that in pattern II.
When the partial charge of the hydrophilic particles is increased to
$Q=0.8$ \emph{e}, the water induced PMF is positive for almost all
the patterns except for pattern I and pattern II in which the water
induced force is attractive in a small range of inter-plate distance
with a minimum at about $d=0.64$ nm (see Fig.~\ref{fig:hetero_5}
(b)).
This corresponds to the cavitation or partial dewetting when two
plates approach toward each other. The water induced repulsive force
in the gap region for pattern V is very large in comparison with that for
the case of $Q=0.4$ \emph{e}. This means that it is very hard to
remove water molecules from the inter plate region with strong
hydrophilic particles.

The solvent induced free energy of interaction (or the solvent induced
part of the PMF) between two plates in the Janus interface as a
function of their separation is shown in Figure~\ref{fig:hetero_5}
(c) for different partial charges on the hydrophilic plate. For both
partial charges ($Q=0.8$ or $0.4$ \emph{e}), the water induced
interactions between the plates are attractive even though the Janus
interface with $Q=0.8$ \emph{e} does not exhibit dewetting while the
interface with $Q=0.4$ \emph{e} does.
Nevertheless, for $Q=0.8$ \emph{e} the magnitude of the attractive interaction, as well as the shape of the curve which exhibits a solvent separated minimum and a barrier to remove this solvent layer, corresponds to an absence of dewetting.
The difference in the solvent
induced PMF for the Janus case and for the case with purely
hydrophobic plates ($Q=0.0$ \emph{e}) (which also shows dewetting)
increases significantly as the partial charge on the hydrophilic
plate increases.  This difference can be as large as ~240 kJ/mol
when $Q$ is $0.8$ \emph{e} relative to that for $Q=0.0$ \emph{e}.

\section{Discussion and Conclusions}
\label{sec:conclusions}

Our previous studies of proteins~\cite{Hua_jpcb07} indicated that
although large matched and connected hydrophobic areas are
correlated with a dewetting transition between two domains or
oligomers, they are not sufficient to predict it. In this work, we
aimed to determine a relationship between the magnitude of surface
hydrophobicity and the spatial distribution of hydrophobic and
hydrophilic domains on the surface. We determined the potential of
mean force and through it the strength of the solvent induced
interaction between two parallel identical amphiphilic plates, which
should be regarded as an idealized model that might serve as a
metaphor for protein inter-domain, or inter-oligomer interactions.
The incorporation of hydrophilic particles in our model system was
performed to mimic the effect of charged and polar side chains on
the properties of the interface. We represented hydrophilic
particles in two ways. The first is as particles with non-zero
partial charges ($\pm Q$), and the second is by LJ particles with a
large well-depth ($\epsilon$), significantly larger than that for
the hydrophobic particles. Since the number of hydrophobic and
hydrophilic particles is constant (and the alignment of the
different type of particles on the two opposing surfaces is
in-registry, see Section~\ref{sec:methods}), the differences in the
behavior of water in the gap must arise from the different spatial
distributions of the hydrophobic and hydrophilic particles on the
plates (at the given strength of water-particle interactions).

We examined five different hydrophobic/hydrophilic particle
distributions on the amphiphilic plates. Our results show that there
are qualitative and quantitative differences in the behavior of the
water for these different distributions. This is manifested by the
existence or absence of a dewetting transition and by attractive or
repulsive solvent induced interactions. Since the existence of a
dewetting transition is sensitive to the strength of the
solute-solvent attractions~\cite{Zhou_science04}, we also studied
how weak and strong hydrophilic particles affect the solvation of
the inter-plate region for different plate patterns. In the case of
weak hydrophilic particles ($Q$=0.4 \emph{e}), most patterns exhibit
a dewetting transition and the observed critical distance for
dewetting, $D_c$, varies for different patterns. The simulation
results show a linear correlation between $D_c$ and a proposed order
parameter that describes the spatial arrangement of the particles on
the plates. This order parameter, the average number of hydrophobic
nearest neighbors ($N_{nn}$), is a measure for the degree of
cooperativity for an assembly of hydrophobic particles on a surface.
For example, for a pattern with a large cluster of hydrophobic
particles at the center of the plates (pattern I), $N_{nn}$ and
$D_c$ are both larger than for any other pattern. On the other hand,
for a pattern where the hydrophobic/hydrophilic particles are
uniformly distributed (pattern V), $N_{nn}$ and $D_c$ are both found
to be smaller than for any other pattern. In this case, no dewetting
transition occurs. Our proposed parameter, $N_{nn}$, seems to be
able to discriminate between the different patterns and correlates
highly with the critical distance for dewetting. For amphiphilic
plates with strongly hydrophilic particles ($Q$= 0.8 {\it e}), the
dewetting transition is suppressed in most patterns. However, the
extent of water depletion in each pattern at the same plate-plate
distance is linearly proportional to $N_{nn}$. Thus, $N_{nn}$ is a
good estimation of overall surface hydrophobicity. In addition,
stepwise cavitations are found in systems with large hydrophobic
domains in regions defined by the clustered hydrophobic domains. We
found that the minimum area of the hydrophobic domain necessary to
induce an adjacent cavity is ~1.04$\times$1.04 nm$^2$. Of course,
the differences in the hydration of the inter-plate region
corresponding to different patterns might lessen or disappear
entirely if the hydrophilicity is made stronger. We did not
investigate how the dewetting transition responds to misaligning the
plates, but we expect it to be sensitive to their relative
orientations. However we did investigate the behavior of water
between two plates forming a Janus interface\cite{granick1_2002}
consisting of one hydrophobic and one hydrophilic plate and found
that the result depends on the polarity of the hydrophilic plate
with the critical distance for dewetting being inversely correlated
with the strength of hydrophilicity of the plate. The drying
transition disappears entirely when the charges on the hydrophilic
plate are sufficiently large ($Q=0.8$ \emph{e}). Our observation is
consistent with an intriguing experiment performed by Granick and
coworkers who investigated the hydrophobicity of a Janus
interface.\cite{granick1_2002}. They found that the hydrophobic
surface prevents macroscopic drying or cavitation of the liquid.
This allowed them to focus on more intrinsic local properties of
interfacial water near extended hydrophobic and hydrophilic
surfaces, and to compare and contrast water behavior in the
different regions. Shear deformations produced by moving the
hydrophobic surface resulted in very large noisy fluctuations
consistent with the picture of damped capillary waves at the
hydrophobic surface arising from partial dewetting. Film-spanning
fluctuations that might lead to macroscopic dewetting between
hydrophobic surfaces were suppressed by pinning of water at the
hydrophilic wall. Our simulations give evidence that the critical
distance $D_c$ of dewetting for the Janus interface can be predicted
based on the simple macroscopic theory according to which it is the
the average of $D_c$'s for two pure hydrophobic plates (same as the
hydrophobic plate in the Janus interface) and two pure hydrophilic
plates (same as the hydrophilic plate in the Janus interface).

In addition we also investigated the effect of the different patterns
on the strength of the inter-plate interaction.
This was done by
calculating the PMF between the plates. In analogy to solvent induced
interactions between hydrophobic particles, we found that for a
pattern with a large hydrophobic cluster, the inter-plate
water-induced interaction is attractive (qualitatively, similar to the
solvent induced interactions between homogeneous hydrophobic
particles). However, for a pattern where the hydrophobic/hydrophilic
particles are uniformly distributed, the solvent induced interaction
is repulsive. This effect is substantial; the difference in the free
energy change for the association process between these two patterns
can be as large as $\sim$230 kJ/mol. Physically, this qualitative
difference in the induced potential of mean force can be attributed to
the fact that for one pattern, it is easier, (negative induced PMF),
on average, to strip off a water molecule from the plate interface
than from around another water molecule in the bulk, while for another
pattern it is harder (positive induced PMF).  The solvent induced
interaction between two plates which form a Janus interface is
attractive (but less so for larger partial charges) even when the
partial charges on the hydrophilic plate are large enough to suppress
dewetting between plates.  The free energy barrier for dewetting for
the strong hydrophilic system ($Q=0.8$ \emph{e}) reflects the free
energy cost of stripping out the layer of water bound to the strongly
hydrophilic surface as the plates approach each other. The PMFs of the
Janus interface which display dewetting ($Q=0.4$ \emph{e}) is
qualitatively similar to other systems that exhibit drying.  However,
for the Janus interface with $Q=0.8$ \emph{e}, where the drying
transition is absent, the curve of the PMF is different in shape and
displays a minimum and a barrier that are associated with a
solvent-separated layer.

It is known that in order to describe stable native structure of known
protein folds in a united (amino-acid) residue description, pairwise
additive interactions are insufficient
\cite{Vendruscolo_pro00,Tobi_pro00}.  It has been
shown~\cite{Chan_pro00,Sorenson_fd98,Plotkin_jcp97,Dill_jbc97} that
only by including information about the many-body interactions, can
one predict protein collapse or folding, similar to what is found in
experiments.  This cooperativity can arise from many groups in the
protein.  In this paper we demonstrated the importance of
cooperativity for hydrophobic particles.  Since non-additivity is
likely to arise in systems with a strong solvent-induced effect, it is
possible that the many-body effect found in proteins originates from
the hydrophobic side-chain residues.

In summary, we used molecular dynamics to study the thermodynamics of
water confined between two amphiphilic plates, and found that
different distributions of hydrophobic and hydrophilic particles on
the plates give rise to qualitatively different large scale water
structures and water induced plate-plate forces. The results are
sensitive to how strongly hydrophobic and hydrophilic the particles
are. Our qualitative conclusions seem to be insensitive to whether the
hydrophilic plates consist of particles which have coulomb
interactions through partial charges or consist of particles that have
strong LJ attractions with the solvent. Since for all patterns
involved in the study of identical plates the number of hydrophobic
and hydrophilic plate particles is the same, the results point to the
breakdown of the Cassie equation and demonstrate that the hydrophobic
interactions are strongly cooperative. We also investigated
plate-plate interactions and the interplate large scale water
structure in the Janus interface between a hydrophobic plate and a
hydrophilic plate and found that when one of the plates is
sufficiently hydrophilic it pins water molecules, eliminating the
fluctuations that lead to drying.

\vspace{1.5cm} \noindent{\Large\bf Acknowledgments} This research was
supported by the National Science Foundation via grant
(NSF-CHE-13401).

\clearpage \pagebreak

\bibliographystyle{jpc}

\clearpage \pagebreak

\begin{table}[float]
\caption[The critical distance, $D_c$, of dewetting transition for
each pattern and its corresponding average number of hydrophobic
nearest neighbor, $N_{nn}$. ]{The critical distance, $D_c$, of
dewetting transition for each pattern and its corresponding average
number of hydrophobic nearest neighbor, $N_{nn}$ (see
Equ.~\ref{eq:nearest_neighbor}).}
\begin{center}
\begin{tabular}{|c|c|c|}
\hline
Pattern & $N_{nn}$ & $D_c$ \\
\hline
I & 3.20 &  1.06 \ \small{(1.04-1.08)} \\
II & 2.88 & 1.02 \  \small{(1.00-1.04)}  \\
III & 2.24 &  0.96 \  \small{(0.92-1.00)} \\
IV & 1.96 & 0.92 \  \small{(0.88-0.96)}  \\
V & 0.00 &  0.68 \ \small{(0.64-0.72)}  \\
\hline
\end{tabular}
\end{center}
\label{tbl:pattern_dewet}
\end{table}

\pagebreak \clearpage

\noindent{\large\bf Figure Captions}

\noindent{\bf Fig.~\ref{fig:hetero_1}}: (a) The hybrid
hydrophobic/hydrophilic plates with five different patterns of
particle distribution, pattern I-V. (b) Janus faced plates: plate1
and plate2. The sphere in cyan represents hydrophobic particle, blue
represent positive charged particle, red represent negative charged
particle.

\noindent{\bf Fig.~\ref{fig:hetero_2}}: (a) The density of water in
the inter-plate region for pattern I-V as a function of inter-plate
distance when $Q=0.4$ \emph{e} for hydrophilic particles. (b) The
critical distance of dewetting transition as a function of the
average number of hydrophobic nearest neighbors for each pattern.
The red line is a linear fitting to the curve with correlation
coefficient $r$ of 0.999.

\noindent{\bf Fig.~\ref{fig:hetero_3}}: (a) The density of water in
the inter-plate region for pattern I-V as a function of inter-plate
distance when $Q=0.8$ \emph{e} for hydrophilic particles. (b) The
view of a slab of water between two plates for pattern II. The water
molecules are superimposed for 150 frames and are classified by
different color based on their location. Water between opposite
hydrophobic domains are represented in green; water between charged
domains are in red and blue; water in the bulk are in silver. (c)
the same as (b) except for pattern III. (d) the same as (b) except
for the system of two plates, each of which is composed by 9
hydrophobic particles arranged in 3$\times$3 square lattice with a
bond length of 0.32 nm.

\noindent{\bf Fig.~\ref{fig:hetero_6}}: (a) The density of the water
molecules confined between two plates which form a Janus interface,
for different hydrophilicity (magnitude of the partial charge $Q$) of
the hydrophilic plate (plate1, see Fig.~\ref{fig:hetero_1} (b)).  Red
corresponds to partial charge of $Q=0.8$ \emph{e}, black for $Q=0.4$
\emph{e}, and green for $Q=0.0$ \emph{e} in which case plate1 is
purely hydrophobic. (b) The density profile of the water molecules
along $z$-axis of the simulation box when the inter-plate distance $d$
is 0.72 nm. The origin $z$=0 nm is the middle point of two centers of
mass of two plates, and the hydrophilic plate is located at the
negative values of the $z$-axis. Only water molecules inside a
rectangular box along the Z-axis of the simulation box with $|X|<=1.1$
nm and $|Y|<=1.1$ nm (in the XY plane) are considered in these
plots. (c) The same as (b) except for $d$=1.24 nm.

\noindent{\bf Fig.~\ref{fig:hetero_4}}: The density of water between
two plates where particles are the same (black, blue) and where
particles are distributed in pattern I (red) and in pattern V
(green). Here hydrophilic particles are represented as neutral LJ
particles with $\epsilon=2.0$ kJ/mol and the same is applied to
hydrophobic except $\epsilon=2.0$ kJ/mol.

\noindent{\bf Fig.~\ref{fig:hetero_5}}: (a) The water induced PMF of
two hybrid plates for pattern I-V as a function of inter-plate
distance when $Q=0.4$ \emph{e} for hydrophilic particles. (b) The
same as (a) except that the charge of hydrophilic particles is
$Q=0.8$ \emph{e} (c) The water induced PMF of two plates which form
Janus interfaces as a function of inter-plate distance with respect
to different partial charge $Q$ of the charged particles on plate1
(see Fig.~\ref{fig:hetero_1} (b)). Black is for partial charge
$Q=0.4$ \emph{e}, red is for $Q=0.8$ \emph{e} and green is for
$Q=0.0$ \emph{e} in which case plate1 is purely hydrophobic and identical to plate2.

\clearpage


\begin{figure}
\begin{center}
\includegraphics[width=5.0in]{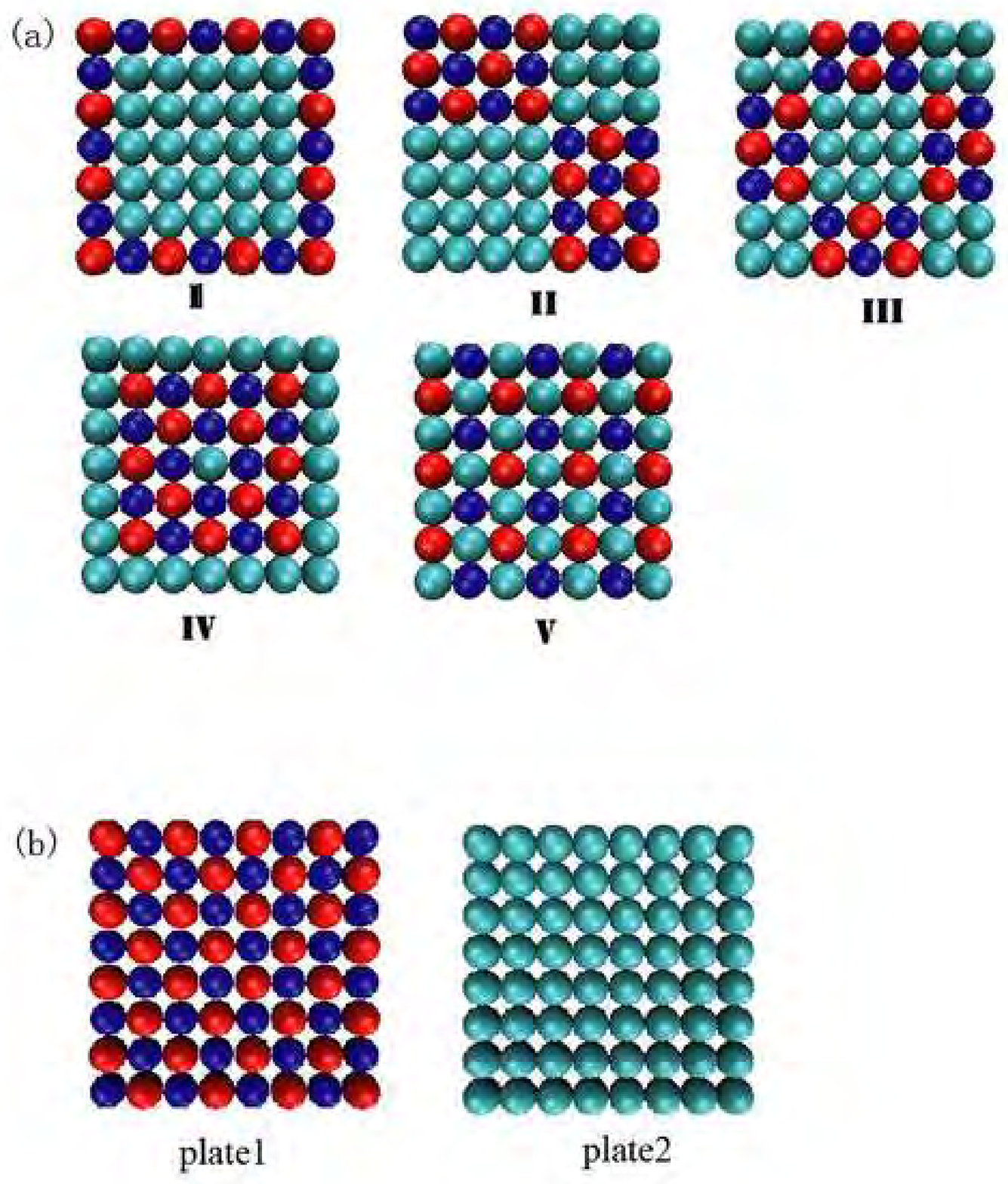}
\caption{\ } \label{fig:hetero_1}
\end{center}
\end{figure}

\begin{figure}
\begin{center}
\includegraphics[width=5.0in]{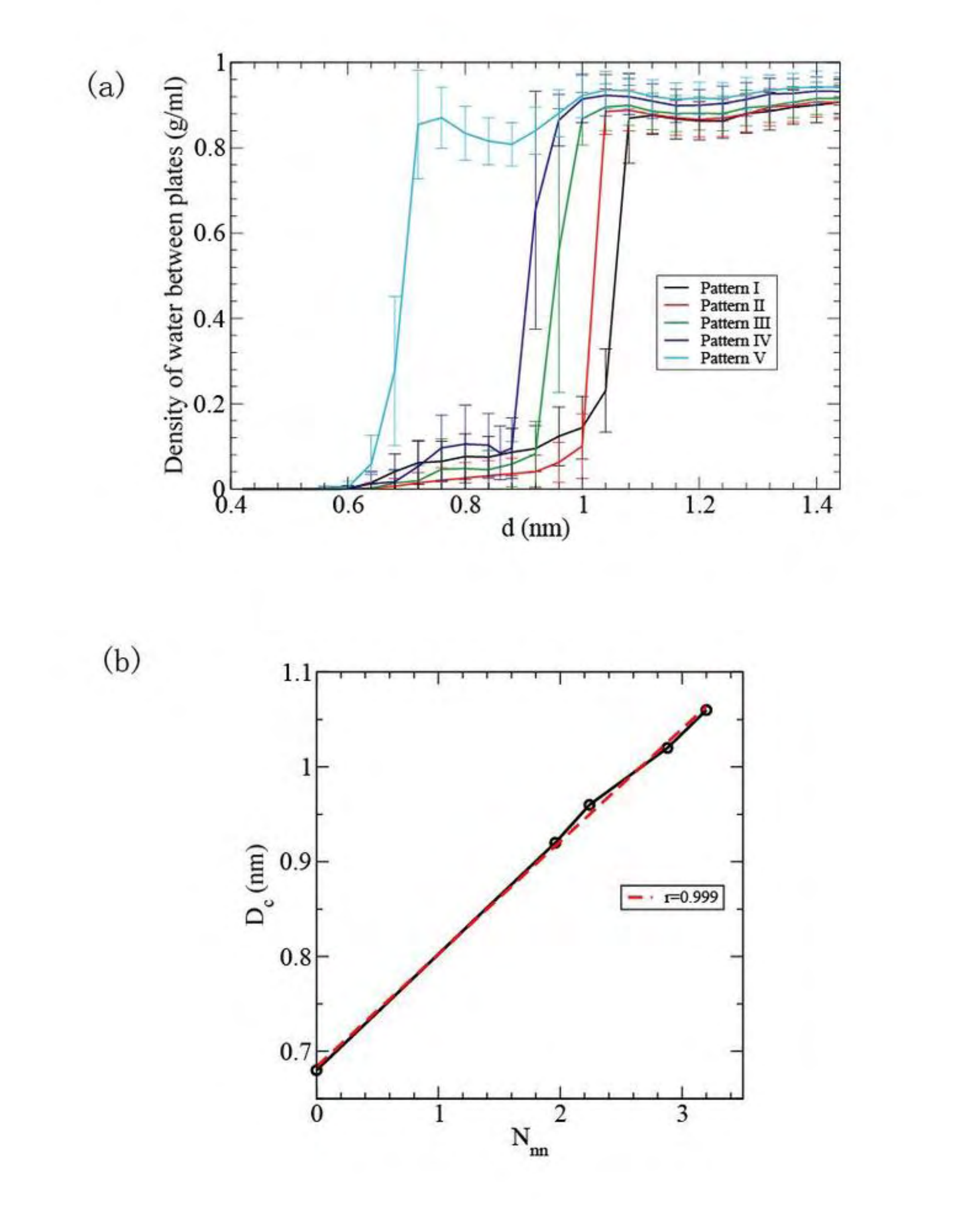}
\caption{\ } \label{fig:hetero_2}
\end{center}
\end{figure}

\begin{figure}
\begin{center}
\includegraphics[width=5.0in]{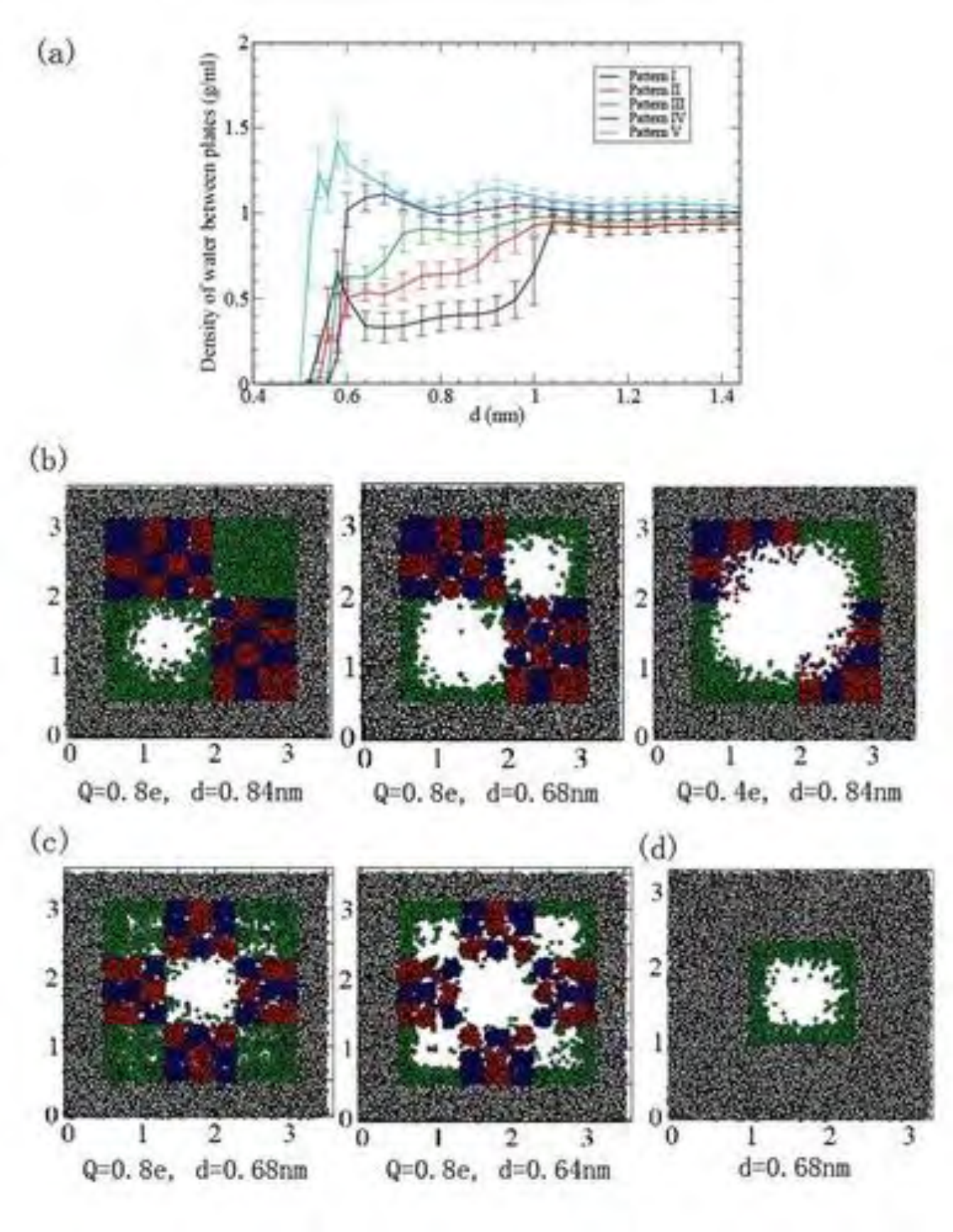}
\caption{\ } \label{fig:hetero_3}
\end{center}
\end{figure}

\begin{figure}
\begin{center}
\includegraphics[width=5.0in]{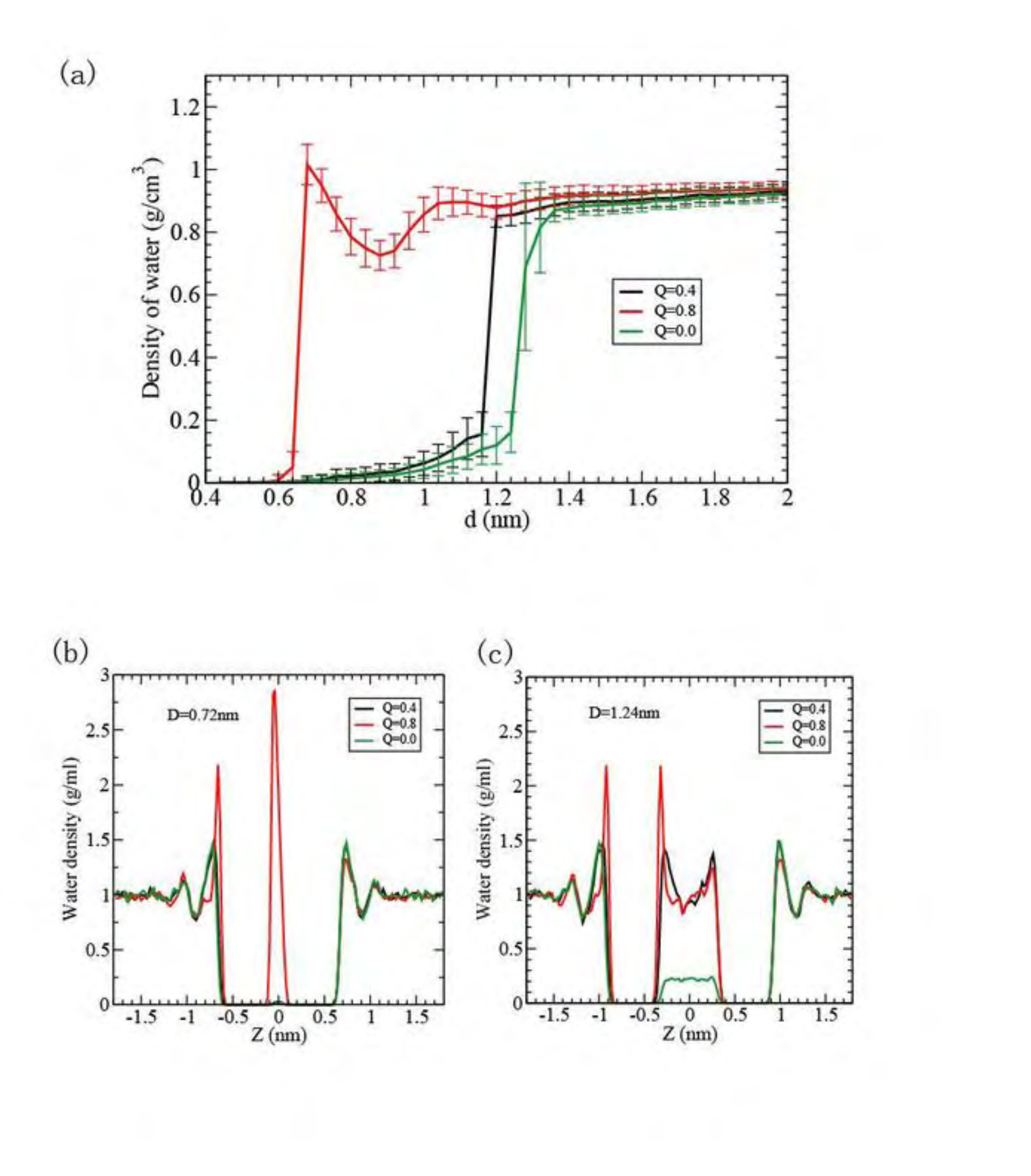}
\caption{\ } \label{fig:hetero_6}
\end{center}
\end{figure}

\begin{figure}
\begin{center}
\includegraphics[width=5.0in]{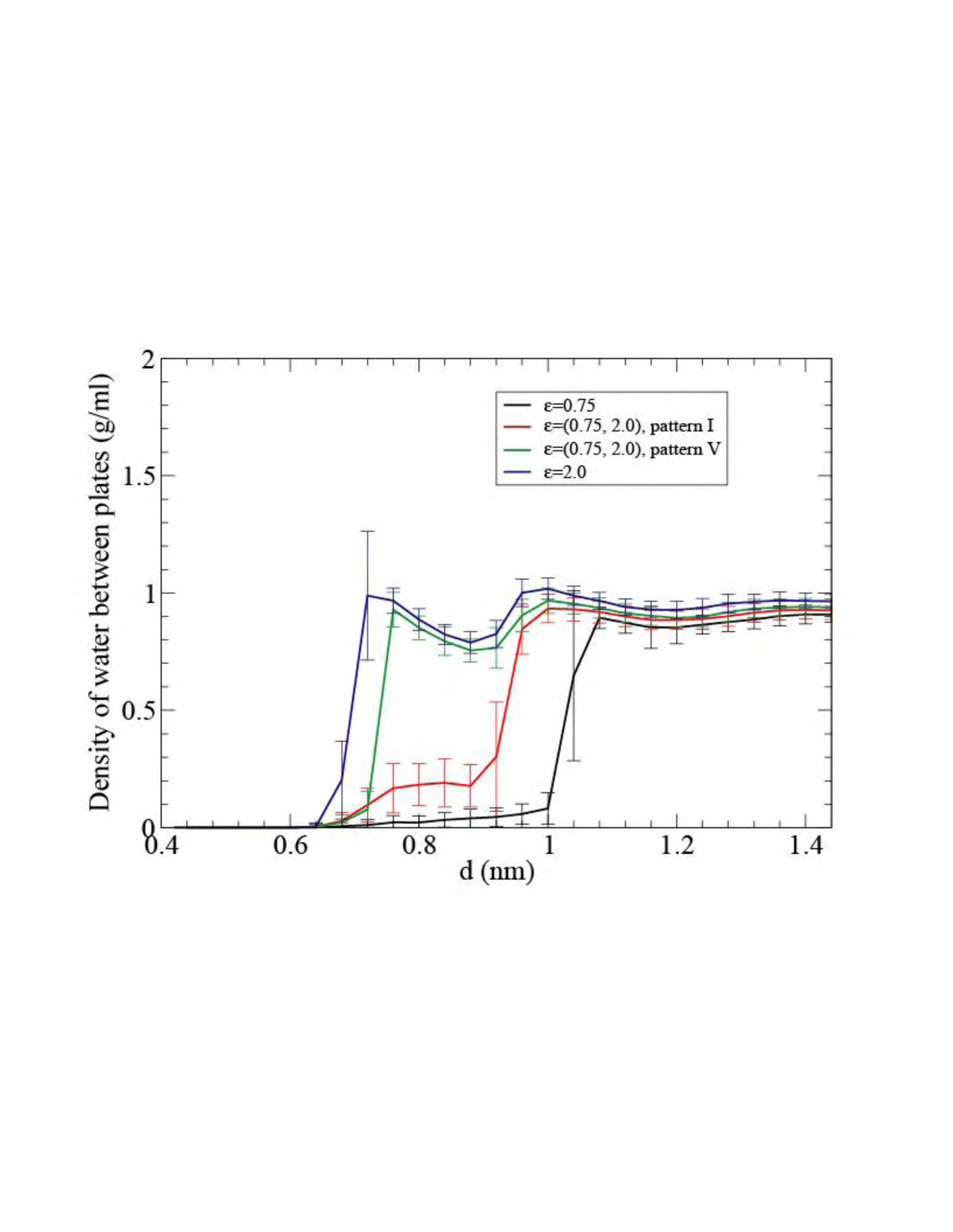}
\caption{\ } \label{fig:hetero_4}
\end{center}
\end{figure}

\begin{figure}
\begin{center}
\includegraphics[width=5.0in]{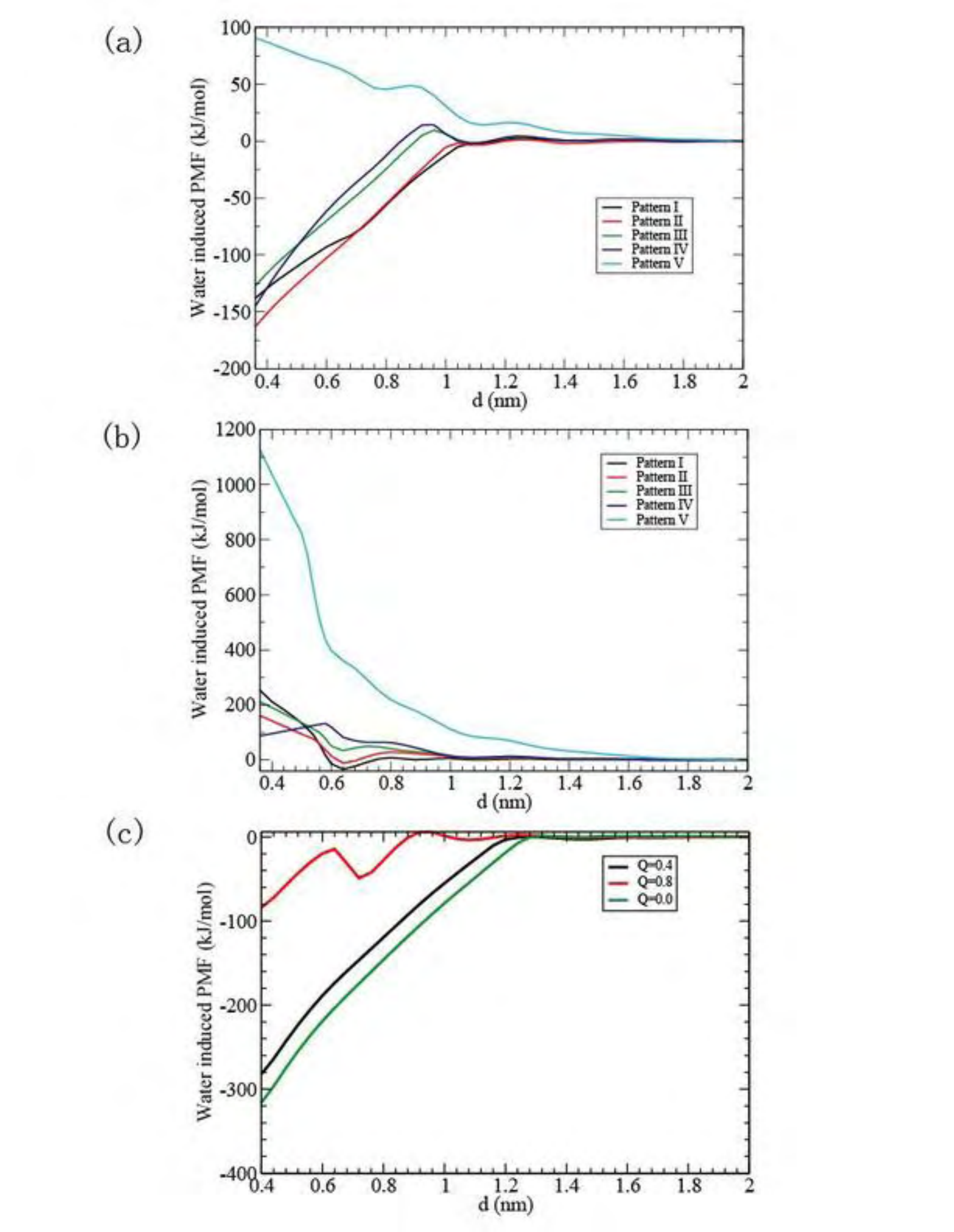}
\caption{\ } \label{fig:hetero_5}
\end{center}
\end{figure}

\end{document}